\documentstyle[12pt,epsf,rotate]{article}

\def\normal{{\bigcirc \!\hskip -5pt n}}

\def\0barra{{\rm O} \!\hskip -3.7pt {\rm l} }

\def\1barra{1\! \hskip -1.1pt {\rm l}}

\title{ Analytical Results of the One-Dimensional \\
           Hubbard Model    in the High Temperature Limit}

\vspace{-0.5cm}

\author{ I.C. Charret$^{(1)}$,  E.V. Corr\^ea Silva$^{(2)}$,
       S.M. de Souza$^{(1)}$, \\
    O. Rojas Santos$^{(3)}$,    M.T. Thomaz$^{(3)}$
                and A.T. Costa Jr. $^{(1)}$   \\
\\
\baselineskip =10pt
{ \small \it $^{(1)}$ Departamento de Ci\^encias Exatas \vspace{-0.2cm}}\\
{ \small \it Universidade Federal de Lavras \vspace{-0.2cm}}\\
{\small \it Caixa Postal  37 \vspace{-0.2cm} }\\
{ \small \it CEP: 37200-000, Lavras, MG,  Brazil }\\
\vspace{-0.2cm}\\
 { \small \it $^{(2)}${\small\it  Instituto de  F\'\i sica} \vspace{-0.2cm}     }\\
 {\small \it Universidade Estadual do Rio de Janeiro } \vspace{-0.2cm}\\
{\small \it R. S\~ao Francisco Xavier n$^{o}$ 524, 3$^{o}$ andar}
\vspace{-0.2cm}\\
{\small \it CEP: 20.550-013, Rio de Janeiro, RJ, Brazil} \\
{ \small \it $^{(3)}$ Instituto de F\'\i sica \vspace{-0.2cm}}\\
{ \small \it Universidade Federal Fluminense  \vspace{-0.2cm}}\\
{\small \it Av. Gal. Milton Tavares de Souza s/n.$\!\!^\circ$,
                  \vspace{-0.2cm} }\\
{ \small \it CEP: 24210-340, Niter\'oi, RJ, Brazil }   \\  }

\date{\today}

\begin{document}

\maketitle

\begin{abstract}

\baselineskip=14pt

We  investigate the grand potential of the one-dimensional
Hubbard model in the high temperature limit,
calculating the  coefficients of the high temperature
expansion ($\beta$-expansion) of this function
 up to order $\beta^4$ by an alternative method.
The results derived are analytical and do not involve any
 perturbation expansion in the hopping constant, being
valid for arbitrary density of electrons
in the one-dimensional model.
 In the half-filled case, we compare our analytical
results for the specific heat and the magnetic
susceptibility, in the high-temperature limit, with the
ones obtained by Beni {\it et al.}
and Takahashi's integral equations, showing that the latter
result does not take
into account
the complete energy spectrum of the one-dimensional
Hubbard model. The exact integral solution by J\"uttner {\it et al}.
is applied
to the determination of the range of validity
of our expansion in $\beta$
in the half-filled case, for several different values of $U$.

\end{abstract}

\vfill

\noindent  PACS numbers: 05., 05.30 -d, 05.30 Fk

\noindent Keywords: One-Dimensional Hubbard Model, Grand
Potential, High Temperature Expansion

\newpage

\baselineskip=18pt

\section {Introduction}

The Hubbard model has been an important candidate
to explain distinct physical phenomena such as itinerant
 magnetism in $4d$-metals\cite{hubbard},  quasi one-dimensional
organic salts (Q1D)\cite{salts} and
superconductivity in high $T_c$ for two-dimensional
 materials\cite{supercondutividade}.

Since the earliest papers on what today is known as the
Hubbard model\cite{hubbard}, perturbation theory has been
used due to the absence of exact solutions in dimensions
higher than one.
For the special case of the one-dimensional
Hubbard model, Lieb and Wu\cite{lieb} applied the
Bethe ansatz in order to get the analytical expression of the
ground state wave function
of the model with
periodic space conditions in the half-filled case. The wave functions
of the excited states of the Bethe ansatz
and their corresponding energies were derived
by  Ovchinnikov\cite{ovchinnikov} from Lieb
and Wu ground state wave function. That was the
situation of the one-dimensional Hubbard model
 at $T = 0$ in the early seventies.

At that time, Takahashi\cite{takahashi} derived an integral
equation for the grand potential of the one-dimensional
Hubbard model, based on the known energy spectrum of the
Bethe ansatz solutions besides  the string hypothesis.
The functions that appear in such integral equation satisfy
an infinite set of coupled equations. At the same time,  Shiba and
Pincus\cite{shiba-pincus} numerically studied
the exact thermodynamics of the Hubbard model
of a  one-dimensional model   with a finite
number of space sites. The longest chain included
six sites with a periodic space boundary condition.
Based on these results, they extended their
conclusions on the behavior of the specific heat and
the magnetic susceptibility to the thermodynamic limit.
Later, Beni {\it et al.}\cite{beni}  applied the standard high temperature
expansion (a perturbation theory in the
hopping constant $t$)  to derive the grand potential
of the one-dimensional Hubbard model up to order
$t^2$. The literature offers many examples of high
temperature expansions of the grand canonical partition
function for the Hubbard model in $d-$dimensions for $ d\ge 2$,
 some of them up to order $\beta^{9}$ ($\beta = \frac{1}{kT}$)
 \cite{kubo,thompson,henderson}. However,
all these works refer to some perturbative  scheme where  one
of the characteristic constants of the model (i.e., its parameters)
must be much
larger than the other ones.

Much more recently,  the  interest  on  theoretical aspects of  the
one-dimensional
Hubbard model returned when Shastry\cite{shastry1, shastry2, shastry3}
proved
its integrability.  A  very interesting approach was developed
where a $d$-dimensional quantum system at finite temperature
is  mapped onto a $d+1$-dimensional classical model. In this
method, the calculation of the grand
potential of the quantum system reduces to obtaining the largest eigenvalue
of the quantum transfer matrix of the $d+1$-dimensional classical model.
This  was successfully applied  to many quantum systems and,
 in particular,
to the one-dimensional Hubbard model by Kl\"umper and Bariev in
'96  for the half-filled case\cite{klumper1}. In '98,
Martins and Ramos\cite{martins} and J\"uttner {\it et
al}.\cite{klumper2} fully performed
 the study of this model at finite temperature.
 In both references,
the largest eigenvalue of the appropriate quantum matrix transfer
was obtained through the
 Bethe {\it ansatz} approach. In reference \cite{klumper2},
 J\"uttner {\it et al}. extended the results obtained by
 Kl\"umper and Bariev
 to any particle density. Differently
from Takahashi's integral solution, the  solution obtained by the quantum
 transfer  matrix  approach includes the solutions with SO(4)
symmetry  for the one-dimensional Hubbard model\cite{outras}.

Charret {\it et al.}\cite{JMP} developed a
method to calculate the analytical expression of the exact
coefficient of the high temperature expansion of the grand
canonical partition function at each order in $\beta$, applying
this method to the calculation of the first three terms of
the grand canonical partition function
 for the one-dimensional Generalized Hubbard
model\cite{physa}. The method is {\em not} based on the
knowledge of the energy spectrum of the one-dimensional
Hubbard model.

In the present paper, we will apply the approach developed in
reference \cite{JMP} to get the coefficients associated to orders
$\beta^3$ and $\beta^4$ of the high temperature expansion
of the grand potential per site of the one-dimensional Hubbard model
subject to a periodic space boundary condition.
Our results are {\it analytical} and do not rest upon any additional
hypothesis on the constants that characterize the model. We should
point out that we do {\it not} perform a perturbative expansion
besides  the high temperature expansion,
that is  we do a $\beta$-expansion of the grand canonical partition
function for any density of electrons. In section 2
we present the grassmannian functions associated to the
model, necessary to the application of the results of reference \cite{JMP}.
In section 3 we present the  coefficients at orders
 $\beta^4$ and $\beta^5$ of the $\beta$-expansion of
the grand canonical partition function of the one-dimensional
Hubbard model. Section 4 is devoted to comparing our results
to the ones known in the literature\cite{takahashi,beni,klumper2}. In
subsection 4.1 we compare
our results to the perturbation expansion
carried out by Beni {\it et al.}\cite{beni}
of the grand potential per site in the hopping term $t$.
  In subsection 4.2 we present a
numerical comparison between Takahashi's results and ours, in the
half-filled case. Finally, in subsection 4.3 we study, for different values
of $U$, the range of validity of our expressions in $\beta$, by
comparing them to the numerical solutions provided by J\"uttner
{\it et al}.   integral equations for the specific heat and
magnetic susceptibility.  Section 5 contains our conclusions.

\section {The One-Dimensional Hubbard Model}

The hamiltonian that describes the one-dimensional Hubbard
model in the presence of an external constant magnetic field
in the $\hat{z}$ direction is\cite{hubbard}:

\begin{eqnarray}
{\bf H} & = &  t \sum_{ i=1}^{\rm N} \sum_{\sigma= -1, 1}
( {\bf a}_{i\sigma}^{\dagger}{\bf a}_{i-1, \sigma} +
{\bf a}_{i\sigma}^{\dagger}{\bf a}_{i+1, \sigma})
+ U \sum_{i=1}^{\rm N} {\bf a}_{i\uparrow}^\dagger
{\bf a}_{i\uparrow} {\bf a}_{i\downarrow}^\dagger
    {\bf a}_{i\downarrow} +  \nonumber \\
%
%
 & & \hspace{3cm}
+ \lambda_B \sum_
{i=1}^{\rm N} \sum_{\sigma=-1, 1} \sigma {\bf a}_{i\sigma}^\dagger
{\bf a}_{i\sigma},  \label {1}
\end{eqnarray}

\noindent where ${\bf a}_{i\sigma}^\dagger$ is the creation
operator of an electron with spin $\sigma$ in the {\it i}-th
site, and ${\bf a}_{i\sigma}$ is the destruction operator of
an electron with spin $\sigma$ in the {\it i}-th site. The
first
term on the r.h.s. of eq.(\ref{1}) is the the hopping term
of the kinetic energy operator with constant $t$.
{\it U} is the strength of the interaction between
electrons in the same site but with different
spins. We have defined $\lambda_B= {1\over 2} g\mu_B B$,
 where {\it g} is the Land\'e's factor, $\mu_B$ is the Bohr's
magneton and $B$ is the constant external magnetic field
in the $\hat{z}$ direction. $N$ is the number of space sites
in the one-dimensional lattice. We use the convention:
$\sigma = \uparrow \equiv 1$ and
$\sigma = \downarrow \equiv -1$.

The periodic boundary condition in space is implemented by imposing
that ${\bf a}_{0 \sigma} \equiv {\bf a}_{N \sigma}$ and
 ${\bf a}_{N+1,\sigma} \equiv {\bf a}_{1 \sigma}$. Therefore,
 the hopping terms $ {\bf a}_{1 \sigma}^\dagger {\bf a}_{0 \sigma}$
 and $ {\bf a}_{N \sigma}^\dagger {\bf a}_{N+1, \sigma}$
 become ${\bf a}_{1 \sigma}^\dagger {\bf a}_{N\sigma}$ and
$ {\bf a}_{N\sigma}^\dagger {\bf a}_{1 \sigma}$ respectively.
We point out that the hamiltonian (\ref{1}) is already in
normal order, and the method by Charret {\it et
al.}\cite{JMP} can be applied directly.

First of all, the high temperature expansion, $\beta \ll 1$,
for the grand canonical partition  function
 ${\cal Z}(\beta; \mu)$ is

\begin{eqnarray}
{\cal Z}(\beta; \mu) & = & {\rm Tr}[ e^{-\beta {\bf K}}] \nonumber \\
%
%
 & = & {\rm Tr}[ 1\!\hskip -1pt{\rm I}] +
\;\;    \sum_{ n=1}^{\infty} {(-\beta)^n \over n!}\;\;
 {\rm Tr}[ {\bf K}^n],
      \label{2}
\end{eqnarray}

\noindent where ${\bf K} = {\bf H} - \mu {\bf N}$,
\ {\bf H} is the hamiltonian of the system,
${\bf N}$ is the total number of electrons operator
 and $\mu$ is the
chemical potential with $\beta = \frac{1}{kT}$; $k$ is
the Boltzmann constant and $T$ is the absolute temperature.

We showed  in reference \cite{JMP} that for  any
self-interacting fermionic quantum system, the
coefficients of the $\beta$-expansion (\ref{2}) can be written as
multivariable Grassmann integrals. For the one-dimensional
($d=1$), these coefficients are

\vskip -0.7cm

\begin{eqnarray}
 {\rm Tr}[{\bf K}^n] &= & \int \prod_{I=1}^{2nN} \, d\eta_I
d\bar{\eta}_I \;\; e^{\sum\limits_{I,J= 1}^{2nN} \bar{\eta}_I\;
A_{IJ} \; \eta_J} \times \nonumber  \\
%
%
 & & \hskip -1cm \times
{\cal K}^{\normal} (\bar{\eta}, \eta; \nu=0)\; {\cal K}^{\normal}
(\bar{\eta}, \eta; \nu=1) \; \cdots \; {\cal K}^{\normal} (\bar{\eta},
\eta; \nu=n-1),  \label {3} 
\end{eqnarray}

\noindent where
$\bar{\eta}$, $\eta$ are Grassmann generators,
${\cal K}^{\normal}$ is the kernel of the ${\bf K}$ operator and
the matrix {\bf A} is given by  \par

\begin{equation}
 {\bf A} = \pmatrix { {\bf A}^{\uparrow \uparrow} & \0barra \cr
 & &   \cr
   \0barra & {\bf A}^{\downarrow \downarrow} \cr }
\label  {4} 
\end{equation}

\noindent so that

\begin{equation}
 {\bf A}^{\uparrow \uparrow} = {\bf A}^{\downarrow \downarrow} =
\pmatrix { \1barra_{N\times N} & - \1barra_{N\times N} &
\0barra_{N\times N} & \cdots & \0barra_{N\times N} \cr & & & & \cr
%
%
\0barra_{N\times N} & \1barra_{N\times N}&
- \1barra_{N\times N} & \cdots & \0barra_{N\times N}\cr
%
%
& & & & \cr \vdots & & & & \vdots \cr
%
%
 \1barra_{N\times N} & \0barra_{N\times N} &
\0barra_{N\times N} & \cdots & \1barra_{N\times N} \cr}.
\label {5}
\end{equation}

\noindent Each matrix  ${\bf A}^{\sigma \sigma}$ has dimension
$ nN \times nN$,
$\1barra_{N\times N}$ and $\0barra_{N\times N}$ being the
identity and null matrices in dimension $N\times N$,
respectively. Here,
{\it N} is the number of space sites and {\it n} is the
power of the $\beta$ term. The matrix {\bf A} is independent of
the particular model under consideration\cite{JMP}.

\vspace{0.5cm}

The kernel of the operator {\bf K}   for the one-dimensional
Hubbard model on the lattice, written in
terms of the Grassmann generators $\bar{\eta}_I$ and $\eta_J$,
is equal to

\begin{eqnarray}
 {\cal K}^{\normal} &(\bar{\eta},\eta;\nu)& =
\sum_{l=1}^N \sum_{\sigma =\pm 1} ( \sigma \lambda_B - \mu) \;
\bar{\eta}_{[{(1-\sigma)\over 2}n +\nu]N + l}\;
\eta_{[{(1-\sigma)\over 2}n +\nu]N + l} +  \nonumber \\
%
%
&& \hspace{-1.5cm} + \sum_{l=1}^N \sum_{\sigma =\pm 1} t [
\bar{\eta}_{[{(1-\sigma)\over 2}n +\nu]N + l}\;
\eta_{[{(1-\sigma)\over 2}n +\nu]N + l+1}+
\bar{\eta}_{[{(1-\sigma)\over 2}n +\nu]N + l}\;
\eta_{[{(1-\sigma)\over 2}n +\nu]N + l-1}] +  \nonumber  \\
%
%
 & & \hspace{-1.5cm} + \sum_{l=1}^N U \bar{\eta}_{(n+\nu)N + l}\;\;
\eta_{(n+\nu)N + l}\;\; \bar{\eta}_{\nu N + l} \; \;
 \eta_{\nu N + l},   \label {6} 
\end{eqnarray}

\noindent where the mapping \cite{JMP}

\vspace{-0.5cm}

\begin{equation}
 \eta_\sigma (x_l, \tau_\nu) \equiv \eta_{[{(1-\sigma)\over 2} n +
\nu]N + l}\hskip 4pt \label {7} 
\end{equation}

\noindent has been used.
The generators $\bar{\eta}_\sigma (x_l, \tau_\nu)$
have an equivalent mapping.  These generators satisfy the
 boundary conditions:

\vspace{0.2cm}

\noindent a) periodic boundary conditions in space:
$\eta_{[{(1-\sigma)\over 2} n + \nu]N + N +1}
 \equiv \eta_{[{(1-\sigma)\over 2} n + \nu]N + 1}$
\hspace{4pt}  \break
and  \hspace{4pt}
$\eta_{[{(1-\sigma)\over 2} n + \nu]N } \equiv
\eta_{[{(1-\sigma)\over 2} n +\nu]N + N}$.

\noindent b) anti-periodic boundary condition in the
temperature ($\nu$):
$\eta_{[{(1-\sigma)\over 2} n +  n]N + l} =
- \eta_{[{(1-\sigma)\over 2} n ]N + l}$, \hspace{4pt}
 for $l= 1, 2, \cdots, N$, and $\sigma= \mp 1$.

In order to write down the terms that contribute
to ${\cal K}^{\normal} (\bar{\eta},\eta;\nu)$ in a simplified way, we define

\begin{eqnarray}
 {\cal E} (\bar{\eta}, \eta; \nu; \sigma) &\equiv &
\sum\limits_{l=1}^N \bar{\eta}_{[{(1-\sigma)\over 2}n +\nu]N + l}\;
\eta_{[{(1-\sigma)\over 2}n +\nu]N + l}\;\; ;
\label {8} 
 \\
%
%
{\cal T}^{\mp} (\bar{\eta}, \eta; \nu; \sigma) & \equiv&
\sum\limits_{l=1}^N \bar{\eta}_{[{(1-\sigma)\over 2}n +\nu]N + l}\;
\eta_{[{(1-\sigma)\over 2}n +\nu]N + l\pm 1} \;\; .
       \label{9} 
\end{eqnarray}

\noindent We also define

\begin{eqnarray}
{\cal E} (\bar{\eta}, \eta; \nu) &\equiv &
\sum\limits_{\sigma=\pm 1} E(\sigma) {\cal E}(\bar{\eta}, \eta; \nu;
\sigma), \label {10} 
      \\
%
%
 {\cal T}^{\mp} (\bar{\eta},
\eta; \nu) &\equiv& \sum\limits_{\sigma = \pm 1} t\, {\cal
T}^{\mp}(\bar{\eta}, \eta; \nu; \sigma),  \label{11} 
\end{eqnarray}

\noindent and

\begin{equation}
 {\cal U}(\bar{\eta}, \eta; \nu)
\equiv U \sum\limits_{l=1}^N \bar{\eta}_{(n+\nu)N + l}\;\;
\eta_{(n+\nu)N + l}\;\; \bar{\eta}_{\nu N + l} \; \; \eta_{\nu N + l},
\label {12} 
\end{equation}

\noindent where $E(\sigma) \equiv \sigma \lambda_B - \mu$. Now,
for the one-dimensional Hubbard model, the grassmannian function
${\cal K}^{\normal} (\bar{\eta},\eta;\nu)$ can be written as
(see eq.(\ref{6}))

\begin{equation}
{\cal K}^{\normal} (\bar{\eta}, \eta; \nu)= {\cal E} (\bar{\eta},
\eta; \nu)+ {\cal T}^{-} (\bar{\eta}, \eta; \nu)+ {\cal T}^{+}
(\bar{\eta}, \eta; \nu) + {\cal U} (\bar{\eta}, \eta; \nu).
\label{13} 
\end{equation}

\vspace{1cm}

\section { The  Coefficients of the $\beta$-Expansion
of the Grand Potential for the One-Dimensional  Hubbard Model}

In reference \cite{physa}, we calculated the  coefficients
of the terms of order $\beta^2$ and $\beta^3$ in
eq.(\ref{2}) for the one-dimensional Hubbard model, for
arbitrary values of the constants  $t$, $U$,
$\mu$ and  $B$ (the external magnetic field).
In this section, the coefficients of
orders $\beta^4$ and $\beta^5$ are calculated.

The evaluation of integrals has been performed by a number of
procedures (computer programs) developed by the authors in the
symbolic system Maple 5.1. This collection of procedures is the
computational
implementation of the method described in reference \cite{JMP}.
We have called this package \footnote{The package can be
downloaded from the site {\it
http:/www.if.uff.br}.} of procedures {\tt GINT}.

The procedure {\tt perm }, contained in the
package, is a useful tool to calculate the independent non-null
terms   that contribute to $Tr[{\bf K}^4]$ and $Tr[{\bf K}^5]$,
and implements
 the symmetries discussed
in reference \cite{physa}. The procedure {\tt gint},
in its turn, calculates the multivariable Grassmann integrals,
taking into account the property of factorization into
sub-graphs (for details, see reference\cite{physa}).

We introduce a simplified notation,

\vspace{-0.4cm}

\begin{eqnarray}
 < {\cal O}_1 (\nu_1) \cdots {\cal O}_m (\nu_m) > &
\equiv & \int \prod_{I=1}^{2nN} \, d\eta_I d\bar{\eta}_I \;\;
e^{\sum\limits_{I,J= 1}^{2nN} \bar{\eta}_I\; A_{I J} \; \eta_J} \;\;
\times   \nonumber \\
 &  &  \hspace{-2cm}
\times \;\; {\cal O}_1
(\bar{\eta},\eta;\nu_1) \cdots {\cal O}_m (\bar{\eta},\eta;\nu_m),
  \label {14} 
\end{eqnarray}

\noindent which let us write the independent terms that
contribute to $Tr[{\bf K}^4]$ in
eq.(\ref{2}) with $n=4$ as

\vspace{-0.5cm}

\begin{eqnarray}
%
Tr[{\bf K}^4] \hspace{-0.5cm} &&  \hspace{-0.3cm}
=<{\cal E} {\cal E} {\cal E}{\cal E}> +
4 <{\cal U} {\cal E} {\cal E} {\cal E}> +
2 <{\cal U} {\cal E} {\cal U} {\cal E}> +
8 < {\cal U} {\cal T}^-  {\cal T}^+  {\cal E}> + \nonumber \\
%
%
& + &4 <{\cal U} {\cal U} {\cal E} {\cal E}> +
4 < {\cal U}  {\cal U} {\cal U} {\cal E}> +
< {\cal U} {\cal U}  {\cal U}  {\cal U}> +
4 < {\cal T}^-  {\cal E}  {\cal T}^+ {\cal E}> + \nonumber \\
%
%
 & + &  8 < {\cal T}^- {\cal U}  {\cal T}^+  {\cal E}> +
4 < {\cal T}^-  {\cal U}  {\cal T}^+  {\cal U}> +
8 < {\cal T}^-  {\cal T}^+  {\cal E}  {\cal E}>  + \nonumber \\
%
%
 &  + & 8 < {\cal T}^-  {\cal T}^+  {\cal U}  {\cal E}> +
 8 < {\cal T}^-  {\cal T}^+  {\cal U}  {\cal U}> +
2 < {\cal T}^-  {\cal T}^+  {\cal T}^-  {\cal T}^+> + \nonumber \\
%
%
 &  + &  4 <{\cal T}^-  {\cal T}^-  {\cal T}^+  {\cal T}^+>.
   \label{15} 
\end{eqnarray}

In order to calculate the terms on the r.h.s. of (\ref{15}),
we need the result of
a set of Grassmann multivariable integrals.
The procedure {\tt gint} is used for obtaining such results. In
reference \cite{physa} we give a lengthy explanation on how
to handle those integrals applying the results of reference
\cite{bjp}, where multivariable Grassmann integrals are
written as co-factors of {\bf A}.

Letting $n=5$, the expression of $Tr[{\bf K}^5]$ outputed by
{\tt perm} is

\vspace{ -0.7cm}

\begin{eqnarray}
Tr[{\bf K}^5]  &= &
 5 < {\cal U}  {\cal E}  {\cal E}  {\cal E}  {\cal E}> +
5 < {\cal U}  {\cal E}  {\cal U}  {\cal E}  {\cal E}> +
5 < {\cal U}  {\cal U}  {\cal E}  {\cal E}  {\cal E}> + \nonumber\\
 & & \hspace{-2cm}
+ 5 < {\cal U}  {\cal U}  {\cal E}  {\cal U}  {\cal E}> +
10 < {\cal U}  {\cal U}  {\cal T}^-  {\cal T}^+  {\cal E}> +
5 < {\cal U}  {\cal U}   {\cal U}  {\cal E}  {\cal E}> +   \nonumber\\
 & & \hspace{-2cm}
+ 5  < {\cal U}  {\cal U}  {\cal U}  {\cal U}  {\cal E}> +
< {\cal U}  {\cal U}  {\cal U}  {\cal U}  {\cal U}> +
< {\cal E}  {\cal E}  {\cal E}  {\cal E}  {\cal E}> + \nonumber\\
  & & \hspace{-2cm}
+ 10 < {\cal T}^-  {\cal T}^+  {\cal T}^-  {\cal T}^+  {\cal E}> +
10 < {\cal T}^-  {\cal T}^+  {\cal T}^-  {\cal T}^+  {\cal U}> +
10 < {\cal T}^-  {\cal U}  {\cal U}  {\cal T}^+  {\cal E}> + \nonumber \\
& & \hspace{-2cm}
+ 10 <{\cal T}^-  {\cal T}^+  {\cal E}  {\cal U}  {\cal E}> +
10 < {\cal T}^-  {\cal T}^+  {\cal E}  {\cal E}  {\cal E}>+
10 < {\cal T}^-  {\cal T}^+  {\cal U}  {\cal E}  {\cal E}> + \nonumber \\
& & \hspace{-2cm}
+10 < {\cal T}^-  {\cal T}^+  {\cal U}  {\cal U}  {\cal E}> +
10 < {\cal T}^-  {\cal T}^+  {\cal U}  {\cal U}  {\cal U}> +
10 < {\cal T}^-  {\cal E}  {\cal T}^+  {\cal U}  {\cal E}> +  \nonumber \\
&  & \hspace{-2cm}
+ 10 < {\cal T}^-  {\cal E}  {\cal T}^+  {\cal E}  {\cal E}> +
10 < {\cal T}^-  {\cal U}  {\cal T}^+  {\cal E}  {\cal E}> +
10 < {\cal T}^-  {\cal U}  {\cal T}^+  {\cal U}  {\cal E}> +  \nonumber \\
& & \hspace{-2cm}
+10 < {\cal T}^-  {\cal U}  {\cal T}^+  {\cal U}  {\cal U}> +
10 < {\cal T}^+  {\cal T}^-  {\cal T}^- {\cal T}^+  {\cal E}> +
10 < {\cal T}^+  {\cal T}^-  {\cal T}^-  {\cal T}^+  {\cal U}> + \nonumber
\\
& & \hspace{-2cm}
+ 10 < {\cal U}  {\cal T}^-  {\cal U}  {\cal T}^+  {\cal E}> +
10 < {\cal U}  {\cal T}^-  {\cal E}  {\cal T}^+  {\cal E}> +
10 < {\cal U}  {\cal T}^-  {\cal T}^+  {\cal E}  {\cal E}> + \nonumber \\
& & \hspace{-2cm}
+ 10 < {\cal U}  {\cal T}^-  {\cal T}^+  {\cal U}  {\cal E}> +
10 < {\cal T}^-  {\cal T}^-  {\cal T}^+  {\cal T}^+  {\cal E}> +
10 < {\cal T}^-  {\cal T}^-  {\cal T}^+  {\cal T}^+  {\cal U}>. \nonumber \\
  &     &  \label{16}
\end{eqnarray}

\vspace{-0.5cm}


\vspace{0.5cm}

The relation between the grand potential   per site ${\cal W}(\beta; \mu)$
and the grand canonical partition function ${\cal Z}(\beta; \mu)$
is

\begin{equation}
 {\cal W}(\beta;\mu) =-  \lim_{ N \rightarrow \infty} \frac{1}{N \beta}
                 \ln {\cal Z}(\beta; \mu). \label {17} 
\end{equation}

From eqs.(\ref{15}), (\ref{16}), (\ref{17}) and the results
(up to order $\beta^3$) derived in reference \cite{physa},
we get the grand potential per site up to order $\beta^4$ for the
one-dimensional Hubbard model,

\begin{eqnarray}
{\cal W}(\beta;\mu) & = & - \Bigl\{ {2\over \beta} \ln 2 +
 \biggl( - {1\over 16} U t^2 \lambda_B^2 -{1\over 16} U^2 t^2 \mu
+ {1\over 1024} U^5 + {1\over 16} U t^2 \mu^2 +
{13 \over 768} U^3 \mu^2  -  \nonumber \\
& - &{1 \over 768} U^3 \lambda_B^2 - {1\over 96} U \lambda_B^4 +
{1\over 96} U \mu^4 - {5\over 768} U^4 \mu -
 {1\over 48} U^2 \mu^3  + {1\over 64} U^3 t^2 \biggr) \beta^4-
   \nonumber  \\
& - &  \biggl( -{1\over 8} t^2 U \mu - {1\over 16} U \mu \lambda_B^2
+ {1\over 96} \mu^4 + {1\over 16} t^4 + {1\over 96} \lambda_B^4
+ { 1\over 1024} U^4 - {1\over 48} U \mu^3
+   \nonumber \\
&  + & {1\over 8} t^2 \mu^2 -
  {1 \over 192} U^3 \mu + {5 \over 96} t^2 U^2 +
{1\over 64} U^2 \lambda_B^2 + {1\over 64} U^2 \mu^2 +
{1 \over 16} \mu^2 \lambda_B^2  +  \nonumber \\
& +&  {1\over 8} t^2 \lambda_B^2 \biggr)\beta^3 +
 \biggl( -{U^3\over 64} + {1\over 16} U \lambda_B^2
- {1\over 16} \mu^2 U + {1\over 16} \mu U^2 \biggr) \beta^2
+   \nonumber \\
 & &  \hskip -0.5cm
+ \biggl( {1\over 4} \mu^2 + {1\over 4}
\lambda_B^2 - {1\over 4} \mu \; U + {t^2 \over 2} + {3\over 32}
U^2 \biggr) \beta - \big(- \mu + {U\over 4} \bigr)  +
{\cal O}(\beta^5) \Bigr\} .
  \label {18}
\end{eqnarray}

\noindent It is important to stress that the
coefficients of the $\beta$-expansion of ${\cal
W}(\beta;\mu)$ are exact for any set of constants
($t$, $U$, $\mu$ and $B$) of this model.
From  eq.(\ref{18}) we can obtain the strong limit
approximation having $U \gg t$, as well as the atomic limit
approximation having $U \ll t$.
No matter how large the values of the constants, the high
temperature expansions still makes sense, provided that those
values are finite. In this
case, the $\beta-$region where expression (\ref{18}) is {\em bona fide}
is diminished.

From expression (\ref{18}), we can derive any physical quantity
for the model at thermal equilibrium at high temperature.
In the following we consider the two quantities:

\vspace{ 0.5cm}

\noindent $i)$ the specific heat at constant length and constant number
of fermions $C_{L} (\beta)$,

\begin{equation}
 C_{L} (\beta) = - k \beta {\partial \over \partial \beta}
\biggl[ \beta^2
{\partial {\cal W}(\beta;\mu) \over \partial \beta}\biggr];
 \label{19}
\end{equation}

\vspace{ 0.5cm}

\noindent $ii)$  the magnetic susceptibility $\chi (\beta)$,

\begin{equation}
 \chi (\beta) = - \big( {1\over 2} g \mu_B\bigr)^2  \;
{\partial ^2 {\cal W}(\beta;\mu)  \over \partial \lambda_B^2 }.
\label{27}  
\end{equation}

\vspace{0.5cm}

In general, the available information is in terms of the density
of electrons in the chain,  instead of the chemical potential. The
density of electrons is given by

\begin{equation}
\frac{N_A}{N}=
- \frac{\partial{\cal W}(\beta;\mu)}{\partial \mu},
  \label{21}
\end{equation}

\noindent where $N_A$ is the number of electrons in the chain and
$\beta$, $\lambda_B$ and all the other constants of the model,
are kept constants in the partial derivative. From eq.(\ref{18})
we get

\begin{eqnarray}
\lefteqn{\rho  \equiv \frac{N_A}{N}  =
 ( - {\displaystyle \frac {1}{16}} \,t^{2}\,U^{2
} + {\displaystyle \frac {1}{8}} \,U\,\mu \,t^{2} +
{\displaystyle \frac {13}{384}} \,\mu \,U^{3} + {\displaystyle
\frac {1}{24}} \,U\,\mu ^{3} - {\displaystyle \frac {5}{768}} \,U
^{4} - {\displaystyle \frac {1}{16}} \,U^{2}\,\mu ^{2})\,\beta ^{4} - }
\nonumber \\
%
%
 & & \mbox{} - ( - {\displaystyle \frac {1}{8}} \,U\,t^{2} -
{\displaystyle \frac {1}{16}} \,U\,\lambda ^{2} + {\displaystyle
\frac {1}{24}} \,\mu ^{3} - {\displaystyle \frac {1}{16}} \,\mu
^{2}\,U + {\displaystyle \frac {1}{4}} \,t^{2}\,\mu  -
{\displaystyle \frac {1}{192}} \,U^{3} + {\displaystyle \frac {1
}{32}} \,\mu \,U^{2} + {\displaystyle \frac {1}{8}} \,\mu \,
\lambda ^{2})\,\beta ^{3} + \nonumber \\
%
%
 & & \mbox{} + ( - {\displaystyle \frac {1}{8}} \,\mu \,U +
{\displaystyle \frac {1}{16}} \,U^{2})\,\beta ^{2} + (
{\displaystyle \frac {1}{2}} \,\mu  - {\displaystyle \frac {1}{4}
} \,U)\,\beta  + 1.
\label{22}
\end{eqnarray}

\section{ Comparison to Previous Results}

In this section, we compare our results derived in
section 3 to the ones presented by Beni {\it et
al.}\cite{beni},  as well as to Takahashi's integral
solution\cite{takahashi} and to the
exact integral
solution derived by  J\"utttner {\it et al}.\cite{ klumper2}. We explore
the results of these
references in the high temperature region, where our
expressions are valid, comparing two derived
thermodynamical quantities, namely, the specific heat and
the magnetic susceptibility,  obtained by each
particular approach.

Throughout this section we have chosen $t=1$; all the remaing constants
in the model are expressed in units of $t$.

\subsection{Comparison to Beni {\it et al.}}

The high temperature expansion has been applied to the calculation of
 the coefficients of
the $\beta$-expansion in eq.(\ref{2}).  Henderson {\it et al}.
\cite{henderson}
calculated the terms  up to $(\beta t)^9$ of this series  for the
single-band Hubbard model in two and three-dimensional lattices.
Bartkowiak {\it et al.}\cite{bartkowiak} calculated the high temperature
expansion of the
extended Hubbard model on the simple cubic lattice up to order
$(\beta t)^6$. On the other hand,   the high  temperature expansion of
the one-dimensional Hubbard model was carried out by
Beni {\it et al}.\cite{beni} up to  order
 $t^2$ only, but to all orders in $\beta$. We believe that it is always
interesting  to compare analytical results, mainly because in
reference \cite{beni}  the perturbation
 expansion is done under the condition $\frac{t}{U} \ll 1$ and
 in the temperature region where $\beta t \ll 1$ whereas our
 results are valid for {\em any}
 ratio  $\frac{t}{U}$.

We compare  our main result (eq.(\ref{18})) for the grand potential per site
 of the one-dimensional Hubbard model  to the
 expansion of eq.(8) of reference \cite{beni}
up to order $\beta^4$.
Even though both calculations allow to handle
the problem with arbitrary density of electrons in the
chain, we consider here the half-filled case ($\rho =1$) only,
when we have $ \mu = \frac{U}{2}$.  The difference between
the expressions derived for the grand potential per site in both
calculations is

\begin{equation}
{\cal W}(\beta) - {\cal W}_{BPH} (\beta) = {\displaystyle
\frac {1}{16}} \,\beta ^{3}\,t^{4},  \label{23}
\end{equation}

\noindent where ${\cal W}_{BPH} (\beta)$ is the grand potential
per site derived from the expressions of Beni {\it et al.}.
Such difference has diverse consequences for distinct physical
quantities. In the case of  the specific heats $C_L (\beta)$
and $C_{BPH} (\beta)$,  the difference between
our result and the one derived  by Beni {\it et al.} is

\begin{equation}
C_L(\beta) - C_{BPH} (\beta) =
\frac {3}{4} \,\beta ^{4}\,t^{4},
\label{24}
\end{equation}

\noindent where $C_{BPH} (\beta)$ is the specific heat
derived from  reference  \cite{beni}. The
difference between these two expressions is independent of $U$,
but the relative error decreases as $U$ increases.

The magnetic susceptibility is obtained from the grand potential per site
 $\cal{W} (\beta)$ through eq.(\ref{27}). We see from eq.(\ref{23})
  that the difference between  $\cal{W} (\beta)$ and
${\cal W}_{BPH} (\beta)$ is independent of the external
magnetic field. Both approaches give the same expression for the magnetic
susceptibility up to order $\beta^4$. It is worth noticing
that the result for this thermodynamic function derived
 in reference \cite{beni} was obtained for
 $\frac{t}{U} \ll 1$, whereas the
 approach presented here allows one to affirm that, up to this
order in $\beta$, it is valid for {\em arbitrary} values of
$\frac{t}{U}$.

In reference \cite{bartkowiak} Bartkowiak {\it et al}. asserted:
``{\it The HTSE method is exact at each order in the inverse temperature
$\beta$}". We have a simple argument to understand this assertion.
In the $\beta$-expansion of the grand canonical partition
function (see eq.(\ref{2})), the $\beta ^n$ term is multiplied by
${\rm Tr}[ {\bf K}^n]$, that is proportional to $t^{2n_1} E^{n_2} U^{n_3}$,
with $2n_1 + n_2 + n_3 = n$. For {\it even} values of {\it n}, the
term of highest  order in the
hopping constant {\it t} that contributes to order  $\beta^n$ is $t^n$.
Actually, this
is the only term to be calculated in order to get the exact coefficient of
the $\beta$-expansion
of  ${\cal Z}(\beta; \mu)$, since all the others terms proportional to
$t^{2n_1}$  ($ n_1 = 0,.., \frac{n}{2}$)  would have been calculated
in lower orders in the $t$-expansion after
the standard   high temperature expansion\cite{pan}. When we include the
term
$(t \beta)^n$ in  the standard high temperature expansion, we lift the
restriction
$\frac{t}{U} \ll 1$, under which perturbation theory is usually done.
When {\it n}
takes odd values,
we do not have the  term $t^n$ since we calculate
traces in expansion (\ref{2}).
Therefore all the coefficients of the term $\beta^n$
would have been calculated already
in the terms proportional to $t^{2n_1}$,  ($n_1 = 0, 1, ...,
\frac{n-1}{2}$).
We only need
to collect all the terms of the form $t^{2n_1} E^{n_2} U^{n_3}$, under the
condition $2n_1 + n_2 + n_3 = n$ and drop the restriction $\frac{t}{U} \ll
1$.

This analysis is coherent with result (\ref{23}), where only the term of
order $\beta^4$
in ${\cal W}(\beta; \mu)$ yielded a
correction to the expansion obtained by
Beni {\it et al.}\cite{beni}.

\subsection{Comparison to Takahashi's Integral Solution}

 In reference \cite{takahashi}, Takahashi
presents a set of coupled integral equations of infinite
order for  the grand potential (see eq.(3.5a) in reference
\cite{takahashi}).
We have obtained a numerical solution of Takahashi's equations by
recursively
iterating those equations up to third order. Doing so, we
have obtained the  specific heat and magnetic
susceptibility in the half-filled case.

Such iterated solution showed very good convergence:
after the second or third iterations, the numerical error
obtained for the grand potential was less than 1\%.

We compare our results and those coming from Takahashi's
equations  for two different values of $U$; namely, $U=4$ and
$U=8$ (in units of $t$).
We have $\beta$ ranging over
the interval $[0, 0.1]$. Within this range (and for the values
of $U$ under consideration) our results  are almost
exact (see section 4.3).
 We  compare two physical quantities: the specific heat and the magnetic
  susceptibility.
 The specific heat curves are shown in figure \ref{fig1}, for
 $U=4$ and $U=8$.
  The difference between our results and Takahashi's solutions
is not due to numerical approximations. To make this point clear,
we show in
figure \ref{fig2} that the relative error between the approaches is
larger than
the estimated numerical error. In figure \ref{fig3} we  compare the
results for the magnetic susceptibility. They agree for small
values of $\beta$, but as $\beta$ increases their difference
again can not be explained
 by numerical errors.

We conclude that, since our results can be considered exact in
the given range of $\beta$, something must be missing in
Takahashi's result.
For such small values of $\beta$, both
results should agree perfectly, but that is not the case.

\subsection{Comparison to the Exact Integral Solution by J\"uttner
{\it et al}.}

In references \cite{martins} and \cite{klumper2}, the
thermodynamic properties of the one-dimensional Hubbard model are
fully determined by two independent approaches; one important
point is that solutions with built-in $SO(4)$ symmetry appear in
both, making their results exact. The same applies to reference
\cite{klumper1}


We take the results by J\"uttner {\it et al}.\cite{klumper2} to
discuss the validity of our expressions, since they have managed
to write  the largest eigenvalue of the suitable quantum transfer matrix
as the solution of a few coupled integral equations.

In reference \cite{klumper2} J\"uttner {\it et al}. extended the
results of reference \cite{klumper1} for arbitrary particle
densities, obtaining a new set of coupled integral equations. In
the half-filled case, even though this set is equivalent to the
one presented in \cite{klumper1}, it offers much better numerical
convergence.

We restrict the comparison of our results with the ones derived in reference
\cite{klumper2} to the half-filled case. Our  physical results are obtained
by
differentiating properly the  grand potential  per
site in eq.(\ref{18}). This comparison
can show the range for $\beta$ in which
are suitable in describing the thermodynamics
of the one-dimensional Hubbard model, for different values of $U$
(in units of
$t$). As in the  previous subsections, we focus on the specific heat and
 the magnetic  susceptibility.

For the specific heat (see figure \ref{fig4}) for $U=1$ at
$\beta= 0.3$ the relative error between results is 0.63\%; at
$\beta= 0.4$ it amounts to 2,26\%. For $U=4$ and $\beta=0.2$, the
error is 0,71\%; at $\beta=0.24$ it becomes 2\%. For $U=8$ our
result at $\beta=0.1$ shows an error of 0.15\% and at $\beta=
0.16$ it is 2\%.

For the magnetic susceptibility we get even better results, as
shown in figure \ref{fig5}. For $U=1$ and $\beta=0.3$ the error
is 0.27\%; at $\beta= 0.4$ the it is 0.88\% (still less than 1\%
!). For $U=4$ and $\beta=0.3$ the error is 0.39\%; for
$\beta=0.4$ it is 1.98\%. Finally, for $U=8$ and $\beta=0.2$ the
error is 0.3\% and for $\beta=0.29$ it becomes 2\%.

These comparisons stress that in all studied cases, the precision
of our analytical solution turned out to be far much better than
our initial expectations.

\section{ Conclusions}

The method developed in reference \cite{JMP} can be easily
applied to the one-dimensional Hubbard model,
allowing us to derive  exact analytical coefficients at each
order of the  $\beta$-expansion of the grand potential.
With the help of the procedure {\tt gint}, written in the
symbolic language {\tt Maple 5.1}, the multivariable
Grassmann integrals can be easily calculated. Besides, the
property of factorization of graphs into sub-graphs,
described in reference \cite{physa}, allows us to
reduce the number of integrals to be actually calculated.
Besides, the method does not involve any further approximation
scheme.   Even though the physics for $U>0$ and $U<0$ are
 different, the results of section 3 apply equally well
for both cases.

Beni {\it et al.}\cite{beni} derived a perturbative expansion in
the hopping constant $t$, valid in the temperature range $\beta t
\ll 1$. It is simple to compare our results to theirs, since both
methods yielded analytical results. For the sake of comparison,
we consider the one-dimensional Hubbard model in the half-filled
case. Equation (\ref{23}) shows that the difference between our
result for the grand potential per site and Beni {\it et al}. is
proportional to $t^4 \beta^3$; it gives a correction to the
specific heat but keeps the magnetic susceptibility unchanged.
Our result is compatible with the assertion of 
Bartkowiak {\it et al}.\cite{bartkowiak}
 that the coefficients derived by the
standard high temperature expansion are exact at each order of
$\beta^n$. With our correction to the grand potential per site
derived by Beni {\it et al}., we can drop the condition
$\frac{t}{U} \ll 1$ under which it was derived.

Takahashi's integral solution\cite{takahashi} is derived from the
energy spectrum of the Bethe {\it ansatz} solution of the
one-dimensional Hubbard model plus the so-called string
hypothesis.   In the case of the specific heat, our 
 correction to Takahashi's result   is not associated
with any numerical approximation or error.
We present  figure \ref{fig2} to show  this fact. For the magnetic
susceptibility there is a larger correction to the Takahashi's
solution in the high temperature region. Certainly, these
differences come from the fact that Takahashi's calculation does
not take into account solutions to the one-dimensional Hubbard
model with $SO(4)$ symmetry\cite{outras}.

In order to determine the range of validity in $\beta$ of our
analytical solution for the grand potential per site for
different values of $U$ (in units of $t$), we considered the
curves of the specific heat and magnetic susceptibility for
$U=1$, $U=4$ and $U=8$. In all cases studied the validity of our
expressions are far beyond our initial expectations. For example,
for $U=1$ the error of our result for the magnetic susceptibility
at $\beta=0.4$ is less than $1\%$. The result of the standard
high temperature expansion is not reliable for $U=1$ and the
numerical solution of the coupled integral equations in the
approach of J\"uttner {\it et al}.\cite{klumper2} are very involved, while
our  analytical result is a very good approximation for $U=1$ up to
$\beta=0.4$.

Finally, we should mention
that the present approach opens the possibility of calculating
the first terms of the $\beta$-expansion of the grand canonical
partition function of the Hubbard model in two space dimensions, as
well as of one-dimensional models with inhomogeneities. We believe that
improvements on the present approach will render a valuable tool for
tackling with such problems.

\section { Acknowledgements}

The authors thank J. Florencio Jr. for interesting
discutions.  O.R.S. and M.T.T. are deeply in debt with Andreas Kl\"umper
for discussing reference \cite{klumper2} and making available the program
that
helped us to develop our own. O.R.S. thanks CAPES
for financial support. M.T.T and S.M. de S. thank
CNPq for partial financial support. A.T.C. Jr, I.C.C and
 S.M.de S. thank FAPEMIG for partial financial support.
 M.T.T. also thanks FAPERJ and FINEP for partial financial support.

\vspace{ 1cm}



\begin{figure}
\caption{ Specific heat from Takahashi's equations
 (solid line) and Charret {\it et al.} (dashed line)
 for:  $U=4$ and  $U=8$. }
   \label{fig1}
\end{figure}

\begin{figure}
\caption{ Relative error in the  specific heat obtained 
from two successive approximations
  to Takahashi's equations (dashed lines) and the difference between
 Takahashi's and  Charret {\it et al.} results (solid lines)
 for:  $U=4$ and  $U=8$. }
   \label{fig2}
\end{figure}

\begin{figure}
\caption{ Magnetic susceptibility from Takahashi's equations
(solid line) and Charret {\it et al.} (dashed line)
 for $U=4$.  }
   \label{fig3}
\end{figure}

\begin{figure}
\caption{ Specific heat from J\"uttner {\it et al}.  equations
 (solid line) and Charret {\it et al.} (dashed line)
 for:  $U=1$, $U=4$ and  $U=8$. }
   \label{fig4}
\end{figure}

\begin{figure}
\caption{ Magnetic susceptibility from  J\"uttner {\it et al}. equations
(solid line) and Charret {\it et al.} (dashed line)
 for:  $U=1$, $U=4$ and  $U=8$. }
   \label{fig5}
\end{figure}

\newpage

\end{document}